\documentclass[a4paper,11pt]{article}
\usepackage[left=2.5truecm,right=2truecm]{geometry}
\usepackage[utf8]{inputenc}
\usepackage{textcomp}
\usepackage[utf8]{inputenc}
\usepackage{setspace}
\usepackage{graphicx}
\usepackage{color}
\usepackage{multirow}

\usepackage{amssymb}
\usepackage{amsmath}
\usepackage{amsthm}

\title{Structural properties and Raman spectra of columbite-type NiNb$_{2-x}$V$_x$O$_6$ synthesized under high pressure}
\author{J. P. Pe\~na$^{1,2}$\footnote{jully.pena-pacheco@neel.cnrs.fr}, P. Bouvier$^1$, O. Isnard$^1$}

\begin{document}

\maketitle

\begin{center}
$^1$Universit\'e Grenoble Alpes, Institut N\'eel CNRS, 25 rue des Martyrs, 38042, Grenoble, 
France \\
$^2$Instituto de F\'isica, Universidade Federal do Rio Grande do Sul, Av Bento Gon\c calves 9500, 91501-970 Porto Alegre, 
Brazil\\
\end{center}

\vspace{1.5truecm}

\begin{abstract}
The complete set of structural parameters of the new series of compounds NiNb$_{2-x}$V$_x$O$_6$ ($0\leq x \leq 2$) with the unusual columbite-type structure is presented here.  
In the samples containing vanadium, this crystalline structure was  stabilized 
by synthesis in conditions of high pressure and high temperature. 
Here we report here the first Raman spectrum for the NiV$_2$O$_6$-\textsl{Pbcn}
polymorph and extend the list of the previously observed modes for the
NiNb$_2$O$_6$.
The evolution of  the vibrational Raman spectrum produced  when the vanadium is substituted for niobium along the series is also presented and discussed. 
This evolution is interpreted by taking into account the changes in the local structural environment of the niobium/vanadium atoms and its influence over the nickel-oxygen bonds around them.
The presence of vanadium atoms favors an  increase of the symmetry in the arrangement of oxygen atoms around the nickel-ones; 
in counterpart, the vanadium is in an octahedral environment which is more distorted than that of the niobium. 
Because of these apparently subtle differences,  the homogeneous distribution of vanadium in the solid solution 
NiNb$_{2-x}$V$_x$O$_6$ is not possible.\\


\end{abstract}

\section{Introduction}

The family of oxides with formula AB$_2$O$_6$ is formed by a set of compounds with such a rich variety of physical properties and crystalline structures,
that the only characteristic that remains constant among them is the stoichiometry proportion 1:2:6 for the atoms  ``A'', ``B'' and ``O'', respectively.
In this formula, ``A'' represents a divalent cation of some alkaline-earth or transition metal, 
``B''  represents a pentavalent transition metal of the fifth
group, and ``O'' represents oxygen.
The physical properties are ruled by both the cations A and B, but the crystalline structure is in most  cases defined by the atom in the B position. 
As a consequence,  the subfamilies of AB$_2$O$_6$ compounds are commonly classified as vanadates, niobates, or tantalates,
depending if the B position is occupied by vanadium, niobium or tantalum, respectively.
In general terms, when synthesized at room pressure, the symmetry of the crystalline lattice reduces from tetragonal in the tantalates to orthorhombic in the niobates to even triclinic in the vanadates.
Here we will use the niobate
NiNb$_2$O$_6$ which stabilizes in a columbite type structure (group 60, \textsl{Pbcn})  as reference and starting compound for the synthesis of  our series of NiNb$_{2-x}$V$_x$O$_6$ samples.

\vspace{3truemm}
In the structure of NiNb$_2$O$_6$ both Ni$^{2+}$ and Nb$^{5+}$ cations are surrounded by six oxygen atoms forming an octahedral environment.
The octahedra around the Ni-atoms are slightly distorted, but those around the Nb-ones are so distorted that they are sometimes called ``cuboids''. 
In the unitary cell of the NiNb$_2$O$_6$ the lattice parameters satisfy the relation 
 $a>b>c$.
This cell is formed by layers of Ni and Nb octahedra stacked along the $a-$axis forming an alternated sequence Ni–Nb–Nb–Ni–Nb–Nb–Ni ~\cite{hanawa}.
Along  the \textsl{c}-axis both the octahedra around the Ni and Nb cations  form zigzag edge-sharing chains.
On the \textsl{ab}-plane the Ni$^{2+}$ chains are arranged in an isosceles triangular geometry \cite{heid, sarvezuk2012} leading to a magnetic behavior of low-dimensional character ~\cite{sarvezuk2012} which is one of the most interesting properties of this compound.
Another recently discovered property of the
NiNb$_2$O$_6$ is its  potential as a photocatalyst under visible light irradiation for the efficient production of H$_{2}$ from water splitting ~\cite{ye}.

\vspace{3truemm}
The group of the AB$_2$O$_6$ vanadates  is the richest one in terms of polymorphism. For example, the compounds  
AV$_2$O$_6$ present a monoclinic structure of space group \textsl{C2/m} when A = Mn, Cd, Mg, or Zn \cite{marlon1, baran}; meanwhile, depending on the temperature of synthesis, at room pressure the structure  of CoV$_2$O$_6$ can be monoclinic ($\alpha-$phase) or triclinic ($\gamma-$phase)  \cite{markkula2012, lenertz2012, dreifus2018}, and it is unequivocally triclinic when A = Ni, Cu.
The vanadates with A = Ni, Mg, Co, Zn, Mn and Cd 
were produced in the columbite-type structure in 1973 by using conditions of synthesis of high pressure/high temperature (HPHT) \cite{gondrand1974}; at the time only partial studies on the structure were done.
Recently, new and more powerful experimental techniques make possible the study of
the physical properties of these compounds thus renewing the interest on them.
In 2017 the structural and magnetic properties of the \textsl{Pbcn-}MnV$_2$O$_6$ \cite{marlon2017} were first reported, and it was later found that the magnetic order is suppressed  in the \textsl{Pbcn-}MnNbVO$_6$ \cite{marlon1}.
A study of the Raman-spectrum at different values of externally applied pressure on \textsl{Pbcn-}MgV$_2$O$_6$ was also recently reported in Ref. \cite{lian_ChinPhysB}.
To our best knowledge no other experimental studies in vanadates with the columbite structure, which is rather unusual for these compounds, were reported.
Here we deeply explore both the structure and  vibrational Raman spectrum  of the whole series of samples
NiNb$_{2-x}$V$_x$O$_6$, all of them crystallized in the \textsl{Pbcn} space group.

\vspace{3truemm}
Among the vanadates we will use NiV$_2$O$_6$ as reference compound. At room pressure it
crystallizes in  the  P\={1} space group.  
This structure is formed by edge-connected NiO$_6-$octahedra forming 1D chains along the \textsl{c-}axis~\cite{belaiche}. 
Those chains  are separated from each other by two planes of alternated octahedral and tetrahedral arranges of oxygen atoms around the V$^{5+}$ ions ~\cite{belaiche}.
The P\={1}-NiV$_2$O$_6$ presents an antiferromagnetically ordered phase below $T_N = 16.4$ K, and 
recent studies have proved it as a visible-light-active photoanode for photoelectrochemical (PEC) water oxidation ~\cite{dang2015}.

\vspace{3truemm} 
In general, the vanadium  oxides have a large range of applications
coming from the different physical properties that are observed in these compounds.
Among the enumerated as the most recent applications discovered of vanadium oxides cited in the reference review by Shvets. \textsl{et. al.} \cite{shvets}
are cathode materials for batteries, electro-chromatic systems, supercapacitors, smart windows, optical switching devices and memory elements.
The diversity of physical properties observed in the vanadates is a consequence of the several types of bonds that the different 
valence states of the vanadium make possible.
In this respect, the sensitivity to the local atomic arrangement of Raman spectroscopy make it a specially useful technique  to understand  the 
physical properties of  vanadates.

\section{Methods}

Samples of NiNb$_{2-x}$V$_x$O$_6$ ($0<x<2$) 
were obtained by submitting  stoichiometric mixtures of
the starting powders P\={1}-NiV$_2$O$_6$  and \textsl{Pbcn}-NiNb$_2$O$_6$ to the extreme conditions of HPHT  shown in Table \ref{thermal_treat}. 
All the synthesis were performed in a CONAC-40 press. The pressure was slowly increased/reduced;
the temperature was risen only after the maximal pressure was reached and it was quenched before the pressure started to be reduced.
The sample
NiNb$_2$O$_6$, which is also a precursor for other samples,  was synthesized by mixing stoichiometric quantities of high purity NiO  and Nb$_2$O$_5$; this mix was heated at 1300 \textcelsius~ for 48 h in a muffle-type oven and then left to cool at the natural cooling rate of the furnace.
In order to obtain the NiV$_2$O$_6$, a mixture of high purity Ni-acetate (NiC$_4$H$_{14}$O$_8$) and V$_2$O$_5$ was used. An excess  of 7\% in mass with respect to the stoichiometrically necessary quantity  of V$_2$O$_5$ was used in order to avoid the formation of Ni$_2$V$_2$O$_7$ which was identified as an spurious phase in previous synthesis. The thermal treatment was performed at 620 \textcelsius~ for 48 h.  
No impurities were observed on the X-rays diffraction pattern for the triclinic P\={1}-NiV$_2$O$_6$. The \textsl{Pbcn}-NiV$_2$O$_6$ ($x=2$) was obtained by submitting the pure powder sample of the triclinic phase to the same conditions showed in Table \ref{thermal_treat}.

\begin{table}
\caption{Synthesis parameters to obtain samples of NiNb$_{2-x}$V$_x$O$_6$ ($0<x\leq2$) in the \textsl{Pbcn} phase.}\label{thermal_treat}
\begin{center}
\begin{tabular}{|c|c|c|c|}
\hline
Compound &  P (GPa) & T (\textcelsius) & time at max P and T (min)\\
\hline
NiNb$_{2-x}$V$_x$O$_6$, $1\leq x\leq2$ & 5.9 & 900 & 60\\
\hline
NiNb$_{2-x}$V$_x$O$_6$, $x=0.66$ & 5.9 & 1100 & 20\\
\hline
\end{tabular}
\end{center}
\end{table}

\vspace{3truemm}
X-rays diffraction (XRD) patterns of finely ground powder samples were acquired in a Bruker D8 Endeavor diffractometer with copper anode ($\lambda_\alpha=1.5418$ \AA).
All the measurements were carried out at room temperature in the Bragg-Brentano geometry. 
The diffraction patterns were refined with the Rietveld method by considering pseudo-Voigt shaped peaks. The software used for the whole refinement process was FULLPROF \cite{fullprofRef}. 
Some of the figures showing the crystalline structure were elaborated by using VESTA \cite{Ref_vesta}. 

\vspace{3truemm}
The micro-Raman measurements were carried out at room temperature and pressure by using a custom-built spectrometer with monochromator ACTON Spectra Pro 2750.
This device is equipped with a 1800 mm$^{-1}$ grating blazed at 500 nm and a Pylon eXcelon CCD camera;
a set of three Bragg filters BNF-Optigrate was used in order to reject the excitation of the 514.4 nm (Cobolt Fandango) laser. 
The spectra were recorded in backscattering geometry with a 40x objective used both to focus the incident laser beam and to collect the scattered light.
The spectrometer was calibrated in wavenumber by using the lines 540.056, 585.249, 588.190, 594.483 and 597.553 nm of the Princeton-Instrument Ne-Ar lamp. 

\vspace{3truemm}
The Raman signal of the samples with $0.5<x\leq2$ is rather weak; 
however, as those samples are relatively dark, 
the laser power was set low enough ($W\sim0.6$ mW) to avoid any influence of laser heating on the Raman spectra.
After the optimization of the beam power, the data  were acquired by taking the mean of two frames of 240 s each. 
In the case of the NiNb$_2$O$_6$ (yellow powder) the Raman signal 
is very intense; thus, in order to avoid saturating the detector, the mean to obtain the final spectrum was taken over ten frames of only 10 s each.
The Raman spectra were decomposed with Lorentzian functions using Fityk software (version 1.3.1) \cite{ref_fityk}.

\section{Results and discussion}

\subsection{X-rays diffraction measurements}

\begin{figure}
\centering
\includegraphics[keepaspectratio,width =8.0truecm]{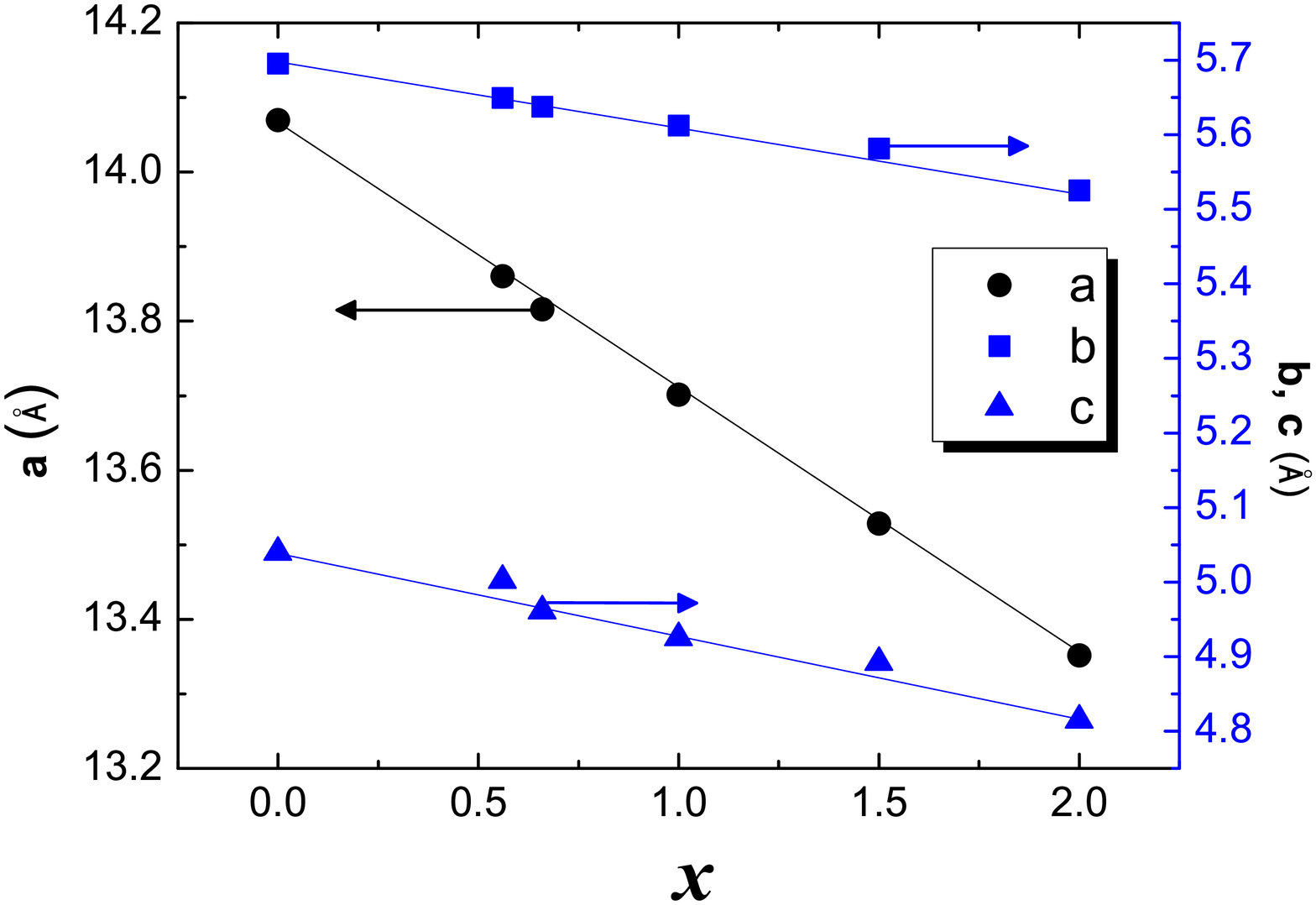}
\includegraphics[keepaspectratio,width =8.0truecm]{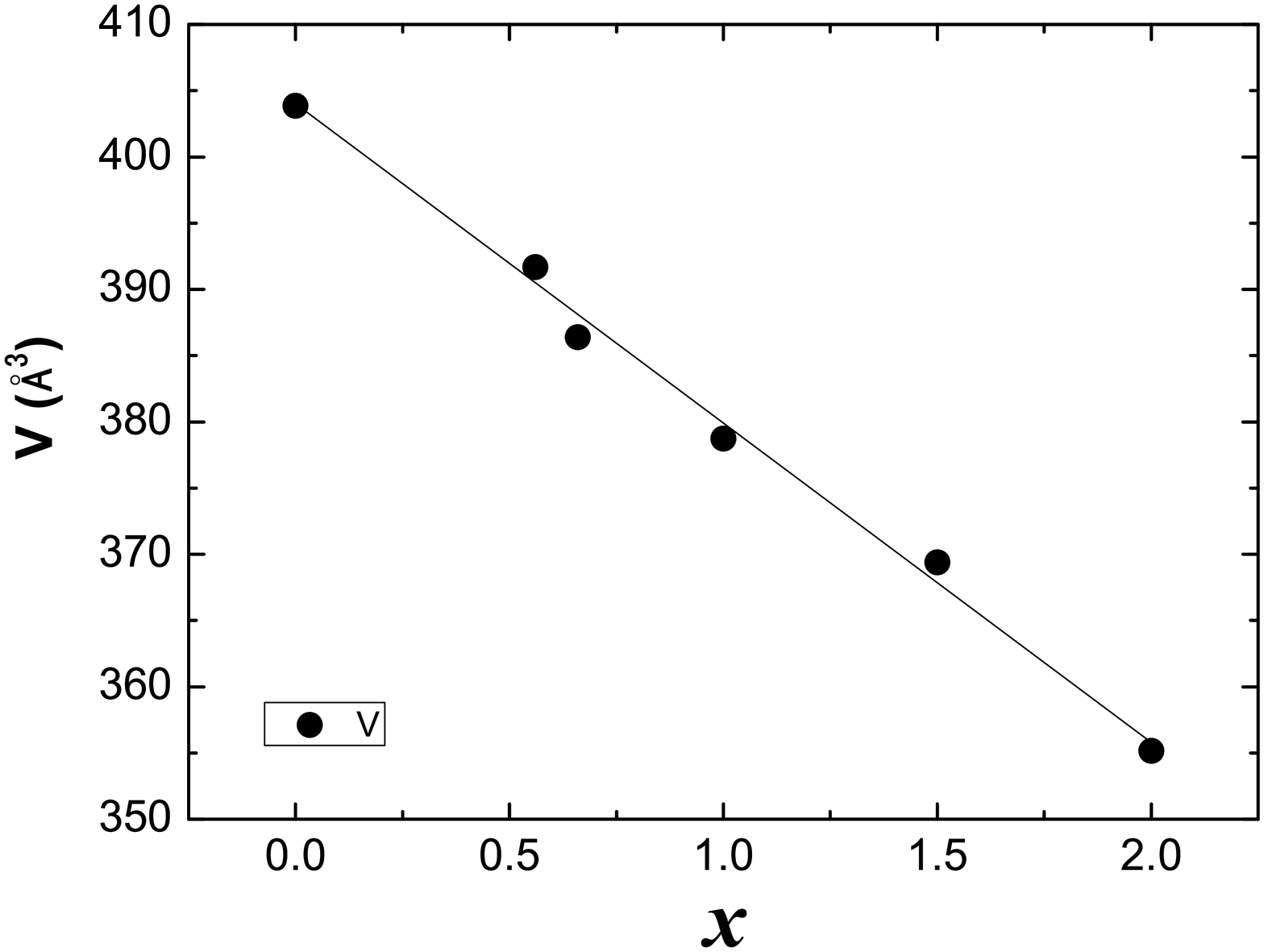}
 \caption{Variation of the structural parameters (left) and the volume of the orthorhombic cell (right) 
  with increasing vanadium content in NiNb$_{2-x}$V$_x$O$_6$. Solid lines are guides to the eyes. } \label{ac_vs_x}
\end{figure}

An investigation of the XRD patterns reveals that the structure of all the synthesized samples of NiNb$_{2-x}$V$_x$O$_6$  retain the orthorhombic symmetry described
by the \textsl{Pbcn-}space group.
The lattice parameters of the \textsl{Pbcn}-NiNb$_{2-x}$V$_x$O$_6$ system shrink 
when the quantity of vanadium increases as shown in Fig. \ref{ac_vs_x}. This effect follows the Vegard's law, and it is expected from the smaller ionic radius of the pentavalent vanadium ions with respect to the niobium ones.
Between the samples $x = 0$ and $x = 2$ the parameters $a$, $b$ and $c$ were noticeably reduced by
4.8\%, 2.7\%,  and 4.1\%, respectively. These values are almost the same as those found
in Ref. \cite{marlon1} for \mbox{A = Mn}, and point to a remarkable anisotropic response of the atomic arrangement to chemical substitution of the B-site in both systems. 
This response seems to be independent of the magnetic nature of the A-ions interactions.

\begin{figure}
\centering
\includegraphics[keepaspectratio,width =10truecm]{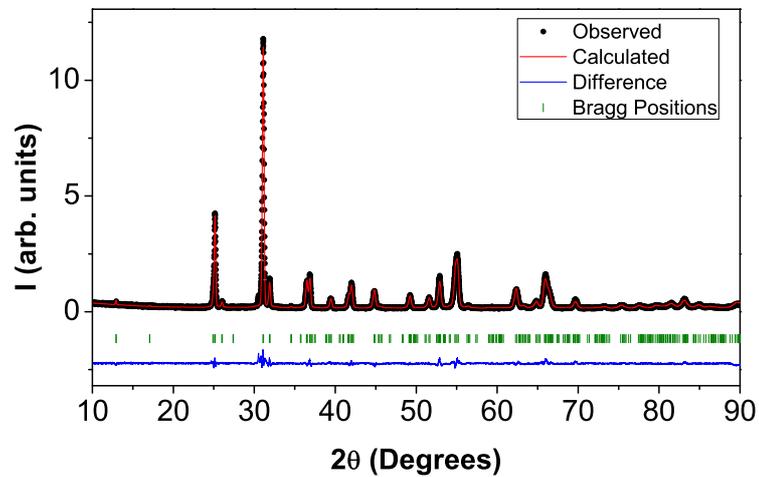}
 \caption{Result of the Rietveld refinement performed at room temperature on the XRD pattern measured for the sample NiNbVO$_6$. The green marks indicate the positions of the expected Bragg peaks from the Pbcn structure.} \label{RietveldJ43}
\end{figure}

\vspace{3truemm}

The complete set of structural parameters for each of the studied samples  and the agreement factors of the Rietveld refinements are reported in Table \ref{structural_param}.
One example of the fitted diffraction pattern is shown in Fig. \ref{RietveldJ43} for the sample  $x=1$.
Our results for sample $x=0$ are in good agreement with those reported in Ref. \cite{sarvezuk2012}.
For all samples, except for that with $x=0.66$, the Rietveld refinement was completed by taking into account only the phase with the desired stoichiometry; thus we infer that our samples are formed by one single structural phase crystallizing in the \textsl{Pbcn} group. Several attempts were made to obtain one sample with one single phase of NiNb$_{1.5}$V$_{0.5}$O$_6$. In all the attempts, it was necessary to include a pure and non-distorted phase of NiNb$_2$O$_6$ in order to complete the refinement; the best result was obtained when we used the most extreme conditions of pressure and temperature  shown on the second row of  Table \ref{thermal_treat}. We got one mixed sample, here called $x=0.66$, where the impurity phase (NiNb$_2$O$_6$) represents 27\% of the total mass of the sample; with this, we calculated that the  main phase, more concentrated in V than what was initially intended, corresponds to the chemical formula NiNb$_{1.34}$V$_{0.66}$O$_6$. 
This stoichiometry is in perfect agreement with the composition dependence of the lattice parameters plotted in Fig. \ref{ac_vs_x}.
The difficulty of obtaining a pure sample of NiNb$_{1.5}$V$_{0.5}$O$_6$ reflects the stability of the orthorhombic structure of the precursor NiNb$_2$O$_6$ which had already been pointed out in Ref. \cite{husson1966} for the isostructural CaNb$_2$O$_6$. Such stability makes it very difficult to distort the NbO$_6$ octahedra with only a modest quantity of vanadium atoms, even under conditions of high temperature and pressure.

\begin{table}
\caption{Cell parameters and atomic positions of the NiNb$_{2-x}$V$_x$O$_6$ (\textsl{Pbcn}-space group) studied samples as derived from Rietveld refinements of  room temperature XRD patterns.}\label{structural_param}
\begin{center}
\begin{tabular}{|c|c|c|c|c|c|c|}
\hline
 \multicolumn{2}{|c|}{$x=$} & 0 & 0.66 & 1 & 1.5 & 2\\
\hline
 {Ni}   & x & 0         & 0         & 0          & 0         & 0 \\
        & y & 0.1612(7) & 0.1613(9)  & 0.1646(6)  & 0.166(1)  & 0.1565(4)\\
        & z & 0.25      & 0.25      & 0.25       & 0.25      & 0.25\\
\hline
 Nb/V   & x & 0.1601(1) & 0.1619(2)    & 0.1628(1) & 0.16477(3)    & 0.1645(2) \\
        & y & 0.3194(2) & 0.3222(3)    & 0.3225(3) & 0.3252(5)     & 0.3242(3)\\
        & z & 0.7502(9) & 0.753(1)    & 0.751(2)  & 0.749(1)      & 0.745(1)\\
\hline
 O1     & x & 0.0843(6) & 0.0852(5)    & 0.0949(3) & 0.1010(6) & 0.0939(5) \\
        & y & 0.383(1)  & 0.372(1)    & 0.395(1)  & 0.389(2)  & 0.412(1)\\
        & z & 0.416(2)  & 0.419(2)    & 0.436(1)  & 0.458(2)  & 0.424(2)\\
\hline
 O2     & x & 0.0775(5) & 0.083(3)   & 0.0805(5) & 0.0815(8)& 0.0882(6) \\
        & y & 0.102(1)  & 0.123(3)    & 0.126(1)  & 0.127(3) & 0.131(1)\\
        & z & 0.919(2)  & 0.917(3)    & 0.908(2)  & 0.903(2) & 0.902(2)\\
\hline
 O3     & x & 0.2628(9) & 0.261(3)    & 0.2542(6)    & 0.251(1) & 0.2552(8) \\
        & y & 0.125(1)  & 0.113(3)    & 0.132(1)     & 0.131(3) & 0.127(2)\\
        & z & 0.582(2)  & 0.584(3)    & 0.584(2)     & 0.583(2) & 0.602(2)\\
\hline 
\multicolumn{2}{|c|}{$a$ (\AA)}  & 14.0229(2)  & 13.8158(4)   & 13.7013(3) & 13.5013(4)   & 13.3518(2)\\
\multicolumn{2}{|c|}{$b$ (\AA)}  & 5.6769(1)   & 5.6375(2)   & 5.6121(1) & 5.5649(2)    & 5.5252(1)\\
\multicolumn{2}{|c|}{$c$ (\AA)}  & 5.0184(1)   & 4.9611(1)   & 4.9253(1) & 4.8639(1)    & 4.8145(1)\\
\multicolumn{2}{|c|}{V (\AA$^3$)}& 399.51(1)    &  386.40(2)  & 378.72(1) & 365.44(2)    & 355.166(8)\\
\hline
\multicolumn{2}{|c|}{$R_P$ (\%)}    & 9.07  & 7.16   & 6.72 & 7.13 & 9.71\\
\multicolumn{2}{|c|}{$R_{WB}$ (\%)} & 11.5  & 9.24   & 9.30 & 9.77 & 12.4\\
\hline
\end{tabular}
\end{center}
\end{table}

\begin{figure}
\centering
\includegraphics[keepaspectratio,width =7truecm]{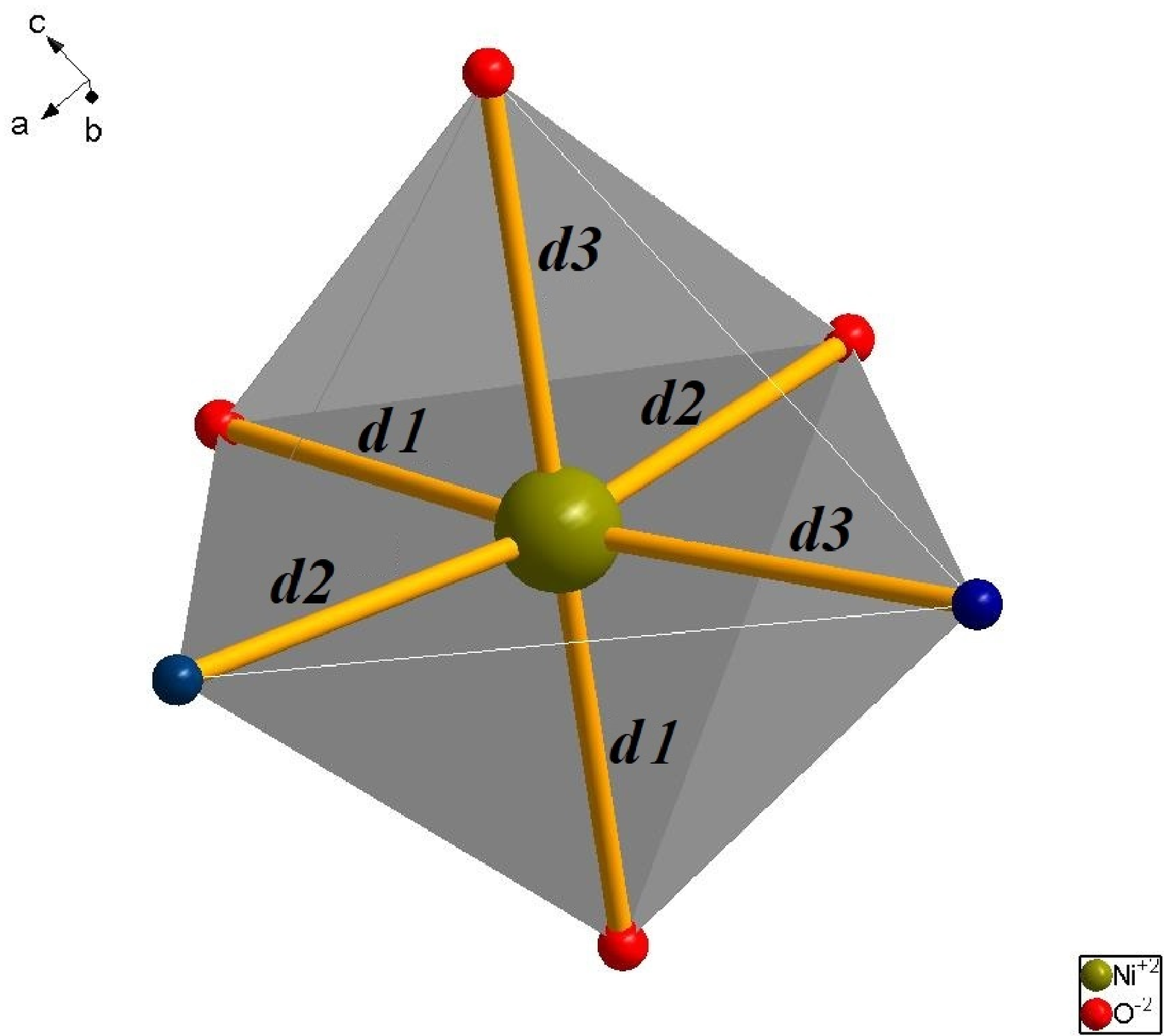}
\includegraphics[keepaspectratio,width =7truecm]{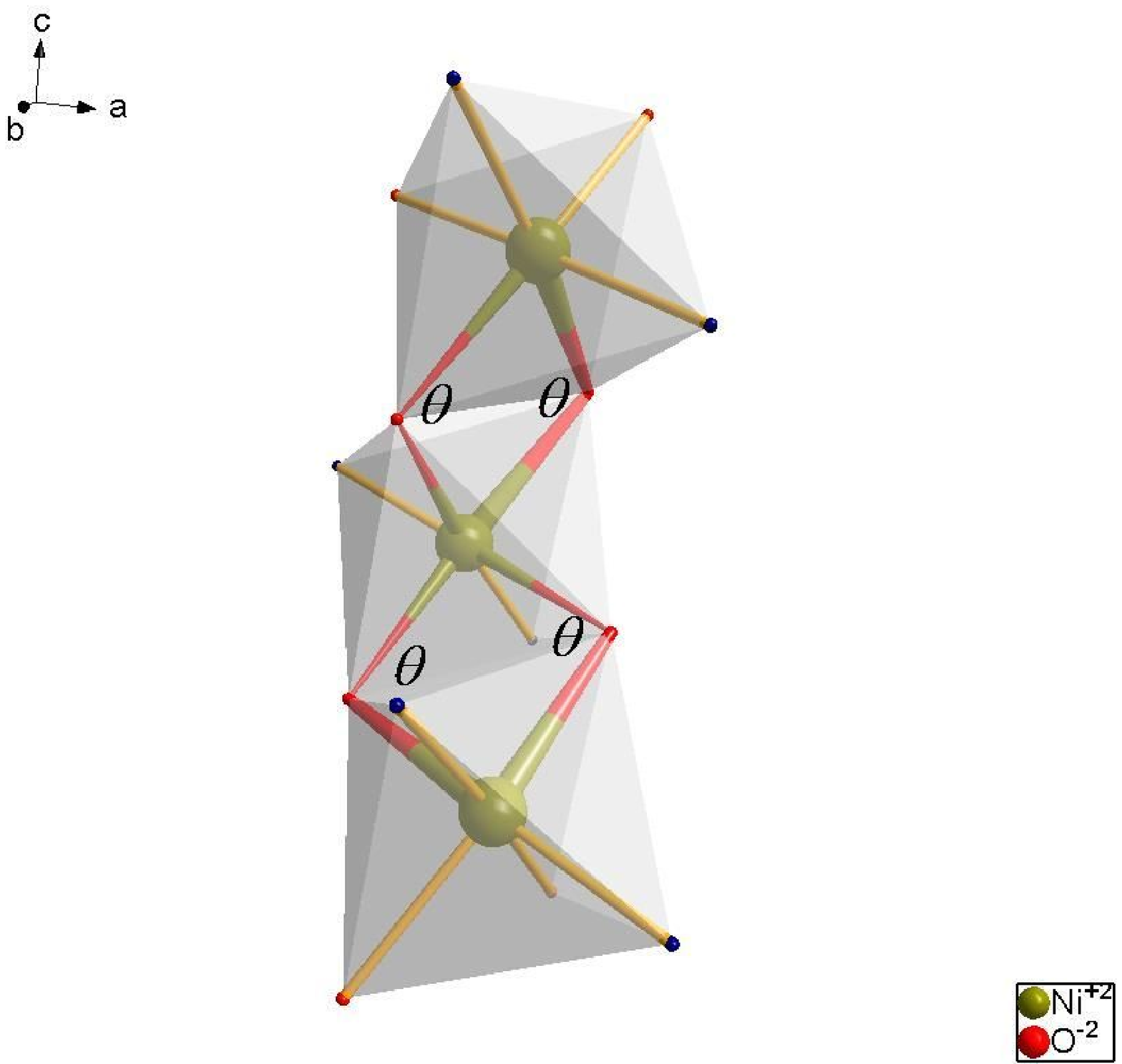}
 \caption{Octahedral environment of the Ni-atoms in the \textsl{Pbcn}-structure (the red and blue spheres represent oxygen-atoms with the non-equivalent crystallographic positions named as O1 and O2, respectively, in Table \ref{structural_param}). At right, the edge-sharing zigzag Ni-chains along the $c-$axis. The angle Ni-O-Ni connecting the octahedra along the chain is signaled. } \label{Ni_octahedra}
\end{figure}

\vspace{3truemm}
The octahedral environment of the Ni-atoms in the columbite structure, and the edge-sharing zigzag chains of Ni-octahedra along the $c-$axis are represented in Fig. \ref{Ni_octahedra}; 
there, the  distances between the central Ni-atom and each  of its six nearest neighbor oxygen atoms are defined. 
These parameters and the distance Ni-Ni along the zigzag chains ($d_{(\textrm{Ni-Ni})c}$) are reported in Table \ref{Ni_distances}.
It is interesting to note here that the three  Ni-O interatomic distances, \textsl{d1}, \textsl{d2}, and \textsl{d3}, increase
with increasing vanadium content whereas the Ni-Ni distance $d_{(\textrm{Ni-Ni})c}$, the cell volume and the unitary cell parameters decrease.
The angle between each of the oxygen atoms on the shared edge and the corresponding adjacent Ni-atoms  along the chain ($\theta$) is also represented in Fig. \ref{Ni_octahedra}, and
its value reported in the last column of Table \ref{Ni_distances}.
By looking at the whole set of data contained in Table \ref{Ni_distances}
one can conclude that the characteristic Ni-chains of the columbite structure suffer considerable alterations when the 
vanadium is substituted for niobium.
Specifically, when the quantity of vanadium increases, 
the Ni-chains are formed by closer Ni-atoms enclosed in larger oxygen-octahedra connected by righter angles.

\begin{table}
\caption{Interatomic distances in \aa{}ngstr\"{o}m between each Ni-atom and the six nearest neighbor oxygen atoms forming its distorted octahedral environment. The last column shows the value of the angle Ni-O-Ni in degrees  along the zigzag chains for each of the samples studied here. The mean value of the interatomic distances Ni-O is represented by the symbol $<d_{\textrm{Ni-O}}>$; all the other symbols are defined in Fig. \ref{Ni_octahedra}.}\label{Ni_distances}
\begin{center}
\begin{tabular}{|c|c|c|c|c|c|c|}
\hline
 $x$  & \textsl{d1} & \textsl{d2} & \textsl{d3} & $<d_{\textrm{Ni-O}}>$ & $d_{(\textrm{Ni-Ni})c}$ & $\theta$\\
\hline
0   & 2.03(1) & 2.02(1) & 1.92(1) & 1.990(4) & 3.106(3) & 100.1(4)\\
\hline
0.66 & 2.139(4) & 2.0266(5) & 1.871(9) & 2.0122(4) & 3.076(4) & 95.1(2) \\
\hline
1.00 & 2.10(1) & 2.07(1) & 2.04(1) & 2,073(4) & 3.081(4) & 95.1(5)\\
\hline
1.50 & 2.11(1) & 2.01(1) & 2.10(1) & 2,078(4) & 3.060(6)& 95.6(5)\\
\hline
2   & 2.110(9) & 2.055(9) & 2.067(9) & 2,077(4) & 2.957(3) & 90.5(4)\\
\hline
\end{tabular}
\end{center}
\end{table}

\begin{figure}
\centering
\includegraphics[keepaspectratio,width =7truecm]{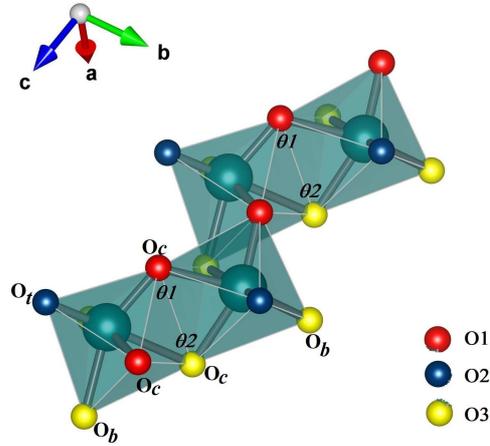}
 \caption{Chain of largely distorted BO$_6$ octahedra along the $c-$axis in the crystalline structure of the NiNb$_{2-x}$V$_x$O$_6$. O$_{t,b,c}\equiv$ oxygen in terminal, bridge, or chain position (see text).} \label{V_chains}
\end{figure}

\vspace{3truemm}
As the Ni-atoms, the B-cations (B = Nb, V) also present an octahedral environment in the \textsl{Pbcn} structure and form chains along the $c-$axis; 
these chains are represented in Fig.  \ref{V_chains}.
The interatomic distances characterizing the octahedral environment around the B-cations, and the distance between two successive B-atoms along the chains ($d_{(\textrm{B-B})c}$) are presented in Table \ref{B_distances}.
It can be seen that the B-octahedra are distorted in such a way that this time there are no pairs of equal distances as it happens for the Ni-octahedra.
Besides, Fig. \ref{V_chains} shows that the two oxygen atoms on the shared edge do not occupy equivalent positions,
and so the angles between the two successive B-atoms and the two O-atoms on the shared edge are different. The values of these angles are presented in Table \ref{B_angles}.

\vspace{3truemm}
For the chains of B-octahedra both interatomic  distances $d_{(\textrm{B-B})c}$ and  $<d_{\textrm{B-O}}>$ decrease upon increasing vanadium content.
Concerning the B-O interatomic distances, \textsl{d1}, \textsl{d2} and \textsl{d3} exhibit a pronounced decrease whereas \textsl{d4} and \textsl{d5} distances are much less sensitive and \textsl{d6} remains almost constant all along the series.
This clearly illustrates the importance of these BO$_6$ octahedra in the \textsl{Pbcn} structure in the context that
these reductions can indeed be interpreted as the origin of the reduction of the unit cell with increasing $x$.
 It is remarkable also that the angles $\theta_1$ and $\theta_2$ evolve in opposite direction with the increasing of $x$,
i.e. $\theta_1$ increases from about 102\textdegree~
to about 109\textdegree~ when $x$ goes from $x=0$ to $x=2$, while $\theta_2$ decreases from about 100\textdegree~ to about 97\textdegree. These results indicate that the BO$_6$ pseudo-octahedra are even more distorted when Nb$^{5+}$ is replaced by V$^{5+}$ in the structure.

\begin{table}
\caption{Interatomic distances in \aa{}ngstr\"{o}m characterizing the B-octahedra and chains of the NiNb$_{2-x}$V$_x$O$_6$ compounds  studied here;  \mbox{$<d_{\textrm{B-O}}>$} is the mean value of the interatomic distances B-O. The subindexes $t$, $b$, or $c$ refer to bonds between the central B-atom and an oxygen in the position terminal, bridge, or chain, respectively (see Fig. \ref{V_chains} and text in the Raman spectroscopy section).}\label{B_distances}
\begin{center}
\begin{tabular}{|c|c|c|c|c|c|c|c|c|}
\hline
 $x$  & \textsl{d1}-O$_b$ & \textsl{d2}-O$_t$ & \textsl{d3}-O$_c$ &  \textsl{d4}-O$_c$ & \textsl{d5}-O$_b$ & \textsl{d6}-O$_c$ &  $<d_{\textrm{B-O}}>$ & $d_{(\textrm{B-B})c}$ \\
\hline
0   & 2.18(1) & 1.89(1) & 1.99(1) & 2.19(1) & 1.99(1) & 2.02(1) & 2.060(4) & 3.241(9)\\
\hline
0.66 & 2.184(9) & 1.760(4) & 1.994(3) & 2.126(3) & 1.98(1) & 1.989(6) & 2.006(1) & 3.189(8)\\
\hline
1.00 & 2.06(1) & 1.74(1) & 1.87(1) &  2.23(1) & 1.85(1) &1.99(2) & 1.958(4) & 3.167(9) \\
\hline
1.50 & 2,06(1) & 1.75(1) & 1.784(5) & 2.204(4) & 1.71(1) & 1.996(6) & 1.916(2) & 3.112(8)\\
\hline
 2   & 1.942(9) & 1.64(1) & 1.77(1) &  2.10(1) & 1.87(1) & 2.04(1) & 1.897(4) & 3.099(8)\\
\hline
\end{tabular}
\end{center}
\end{table}

\begin{table}
\caption{Angles in degrees between the B and O-atoms connecting the BO$_6$-octahedra along the $c-$axis in the NiNb$_{2-x}$V$_x$O$_6$ compounds. The symbols are  represented in Fig. \ref{V_chains}.}\label{B_angles}
\begin{center}
\begin{tabular}{|c|c|c|}
\hline
 $x$  & $\theta1$ & $\theta2$\\
\hline
0   & 101.7(6) & 100.2(6)\\
\hline
0.66 & 99.6(5) & 101.6(4) \\
\hline
1.00 & 108.0(6) & 97.2(7) \\
\hline
1.50 & 111.0(6) & 95.5(4)\\
\hline
 2   & 108.6(6) & 96.7(5)\\
\hline
\end{tabular}
\end{center}
\end{table}

\subsection{Raman spectroscopy}

\begin{figure}
\centering
\includegraphics[keepaspectratio,width =11truecm]{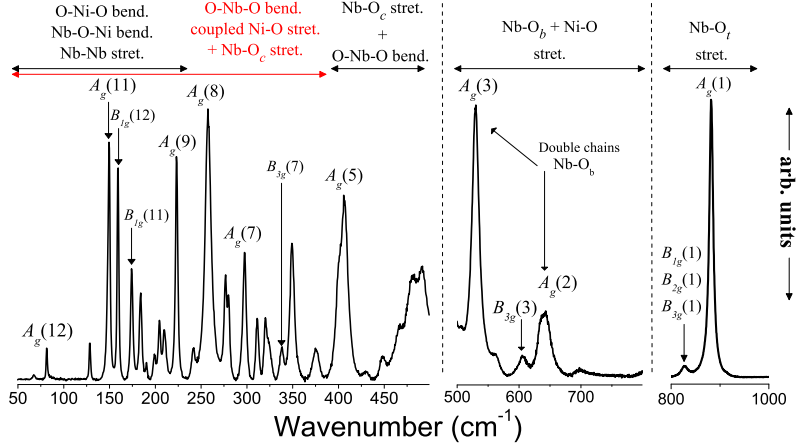}
 \caption{High quality Raman spectrum of the NiNb$_{2}$O$_6$. 
 The three regions of relative wavenumber mentioned in the text are separated by the vertical dashed lines. In each region the vertical axis is normalized to the height of the peak with maximum height  in each region. Stretching and bending are abbreviated as stret. and bend., respectively.} \label{Nb_Raman_spectrum}
\end{figure}

\vspace{3truemm}
The highly resolved Raman spectrum  measured for our sample of NiNb$_2$O$_6$ ($x=0$) is shown in Fig. \ref{Nb_Raman_spectrum}; 
we observed 47 of the 54 expected modes for the structures belonging to the \textsl{Pbcn} space group thus providing significant additional information to the previously reported data on this compound. 
The detailed list of Raman active modes is reported in the Supporting Information.
The spectrum shown in  Fig. \ref{Nb_Raman_spectrum} is representative of the Raman spectrum of all the niobates of the AB$_2$O$_6$ family.
In Refs. \cite{husson1966,  husson_jan77, husson_Ag77} Husson \textsl{et. al.} made a wide theoretical and experimental study of the Raman and infrared modes of vibration of several niobates where the position ``A'' was occupied by Ca, Mg, Mn, Ni, Zn or Cd. 
The authors established that  the Raman spectra of all these niobates are qualitatively equal at wavenumbers higher than \mbox{$\bar{\omega}\sim380$ cm$^{-1}$}. 
The 
region in relative wavenumber larger than $\bar{\omega}>800$ cm$^{-1}$ is dominated by a very intense peak which corresponds to one
 $A_g$ symmetry mode.
This mode is related to the stretching of the Nb-O$_t$ bonds, where O$_t$ denotes the oxygen atoms in the ``terminal or free'' position  which connect the BO$_6$ octahedra with the adjacent chains of NiO$_6$ octahedra.
Between $\bar{\omega}=380$ cm$^{-1}$ and $\bar{\omega}=800$ cm$^{-1}$ the modes 
correspond to the stretching movements inside 
the NbO$_6$ octahedra. 
There the bonds can imply oxygen atoms in the ``bridge'' (O$_b$) and ``chain'' (O$_c$) positions which denote the oxygen atoms connecting two BO$_6$ octahedra belonging to the same chain,
or the oxygen atoms linking two successive BO$_6$ chains, respectively.
Below $\bar{\omega}=380$ cm$^{-1}$  the peaks observed in the spectrum are  due
to modes of bending and torsion of the NbO$_6$ octahedra
involving the angles O-Nb-O. 
In this region, some of those modes are also coupled with displacements of the Ni-atoms. 
Finally, at very low frequencies ($\bar{\omega}<100$ cm$^{-1}$) one observes the modes related to sub-lattices vibrations with respect to each other (for example the harmonic movement of the Ni-atoms with respect to the Nb-ones).

\begin{figure}
\centering
\includegraphics[keepaspectratio,width =10truecm]{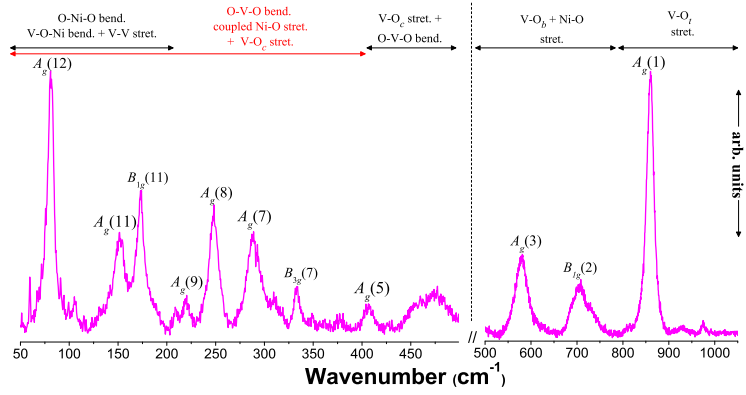}
 \caption{Raman spectrum for the sample  NiV$_2$O$_6$ with the \textsl{Pbcn} structure studied here. The assignement of the modes are presented in Table \ref{Raman_peaks}. The vertical axis is normalized to the height of the peak with maximum height  in each of the two regions separated by the dashed line. Stretching and bending are abbreviated as stret. and bend., respectively.} \label{Raman_x2}
\end{figure}

\vspace{3truemm}
The Raman spectrum for sample NiV$_2$O$_6$ ($x=2$) is shown in Fig. \ref{Raman_x2}.
To our best knowledge, this is the first  reported Raman spectrum of NiV$_2$O$_6$ with the unusual columbite-type structure. 
When compared to each other, the spectrum of sample $x=2$ presents   much wider peaks than that of sample $x=0$; in the region $\bar{\omega}>500$ cm$^{-1}$ the FWHM of individual peaks is around 12 cm$^{-1}$ when $x=0$, while it is around 20 cm$^{-1}$ when $x=2$. 
The origin of this widening can reside in
the increased structural distortion of the VO$_6$ octahedra with respect to the NbO$_6$-ones as already pointed in the previous section. 
Quantitatively, the calculated distortion \cite{fullprofRef, brown1973} ($\Delta= n^{-1} \sum_i^n(\frac{d_i-\overline{d}}{\overline{d}})^2$, where $n$ is the coordination number, and $\overline{d}=<d_{\textrm{B-O}}>$)
of the VO$_6$ octahedra is $\Delta=6.7 \times10^{-3}$, against $\Delta=2.7 \times10^{-3}$ for the NbO$_6$ octahedra. 
As a consequence of the widening of the Raman peaks for the NiV$_2$O$_6$, only a small quantity of modes of vibration  could be identified here for that sample. 
The positions of the peaks that can be identified in  the Raman spectra of the two extreme compounds of the series NiNb$_{2-x}$V$_x$O$_6$ are presented in the second and nineth columns of Table \ref{Raman_peaks}.
As there are no theoretical studies on the normal coordinate analysis of the AB$_2$O$_6$ compounds with the columbite structure when B = V, 
the identification of the symmetry of the modes 
presented in the first column of \mbox{Table \ref{Raman_peaks}}
was done  by comparing our results with those of Raman studies performed in niobates of the AB$_2$O$_6$ family  already reported in the literature   \cite{ husson_jan77, husson_Ag77, huang, husson_jun77}.

\begin{figure}
\centering
\includegraphics[keepaspectratio,width =9truecm]{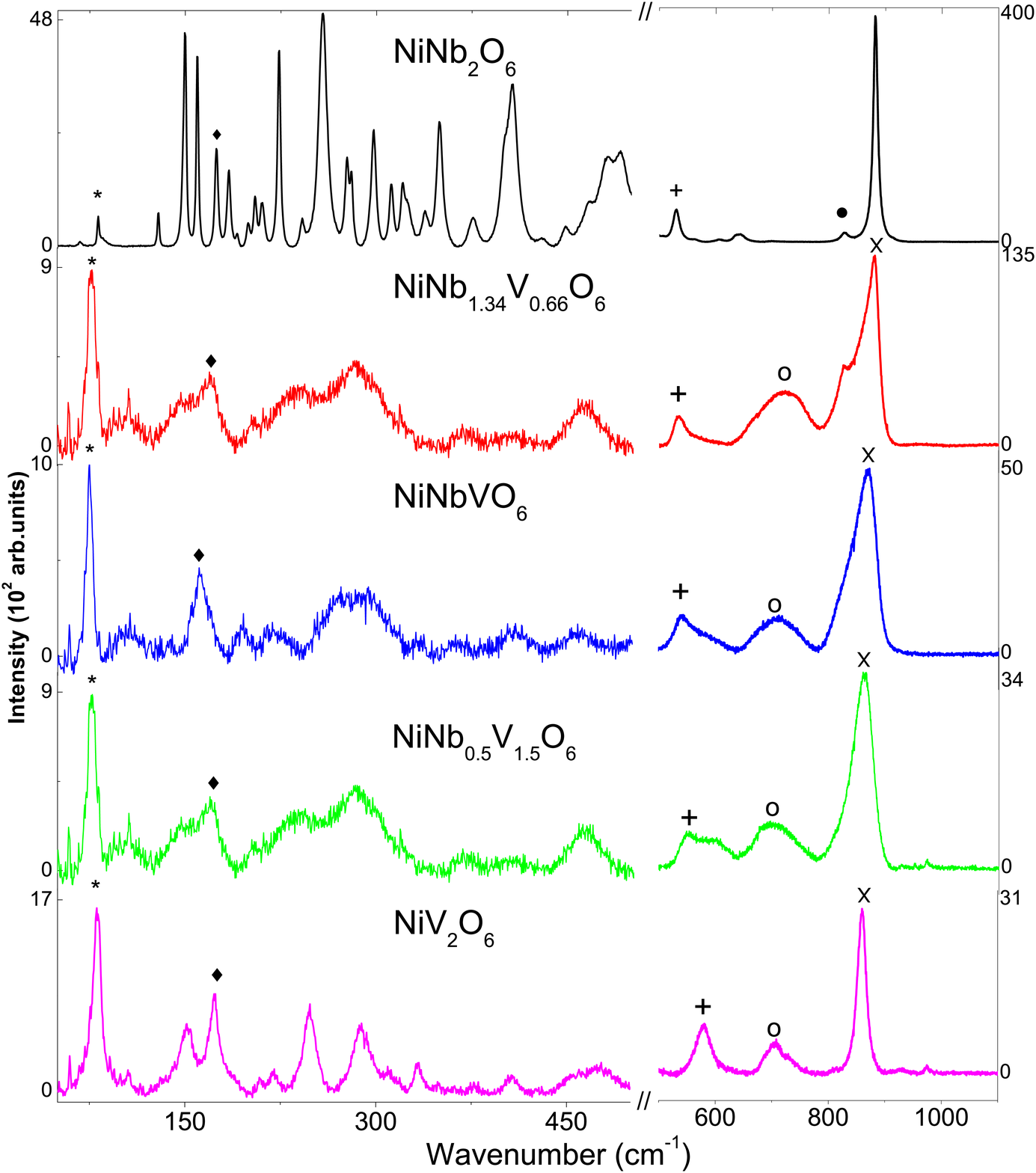}
 \caption{Raman spectra for the complete NiNb$_{2-x}$V$_x$O$_6$ series of samples. Some characteristics common to most of the spectra are marked by symbols. Note that the vertical scale is adapted for each sample and wavenumber domain. } \label{raman_specta}
\end{figure}

Raman spectra acquired for the complete series of NiNb$_{2-x}$V$_x$O$_6$ samples at room temperature and pressure are presented together in Fig. \ref{raman_specta}.
It is observed that, as the widening of the Raman peaks seems to be intrinsic to the  VO$_6$ octahedra, this widening is propagated along the samples with intermediate compositions of Nb and V. 
However, in those samples factors related to the presence of both Nb$^{5+}$ and V$^{5+}$ cations can also
contribute to the enlargement of the FWHM.
The peaks (or sets of peaks) that seem to be common to the Raman spectrum of most of the samples  are signaled by small symbols in \mbox{Fig.  \ref{raman_specta}}. 
As it will be discussed later, in the samples with $0<x<2$ 
the wide peaks correspond to the sum of several Raman peaks that in most cases can not be separated individually.

\vspace{3truemm}
We will analyze the transformation of the Raman spectrum along  the  series  NiNb$_{2-x}$V$_x$O$_6$ 
by dividing the spectrum in three different regions.
The first region is above $\bar{\omega}=800$ cm$^{-1}$ where the most intense peak  of each spectrum is observed; the relative wavenumber where the maximum of that peak appears varies  from \mbox{$\bar{\omega}=882$ cm$^{-1}$}  for the sample  $x=0$  to \mbox{$\bar{\omega}=859$ cm$^{-1}$} for the sample $x=2$.
The second region is between \mbox{$\bar{\omega}=500$ cm$^{-1}$} and \mbox{$\bar{\omega}=800$ cm$^{-1}$}; it 
presents very wide peaks in the samples with $0<x<2$. 
Finally, the third region is the one below $\bar{\omega}=500$ cm$^{-1}$ where the general aspect of the spectrum of the samples with $x>0$ is similar.
In this third region most of the modes are clearly identified only in the two extreme compounds; 
the modes of the sample $x=0$  are observed at slightly higher or very close values of relative wavenumber compared to those in the sample $x=2$. 

\begin{table}
\caption{Raman peaks identified in the spectra of the series NiNb$_{2-x}$V$_x$O$_6$ studied here. The relative wavenumber ($\bar{\omega}$) is given in units of cm$^{-1}$.
For samples with $0<x<2$ the position of the peaks reported is the found from fitting the experimental curves by taking into account the two modes behavior (see text).
The symbol  * indicates that the corresponding mode could not be isolated in the respective Raman spectrum.}\label{Raman_peaks}
\begin{center}
\begin{tabular}{|c|c|c|c|c|c|c|c|c|}
\hline
$x\longrightarrow$  & 0 & \multicolumn{2}{|c|}{0.66} & \multicolumn{2}{|c|}{1} & \multicolumn{2}{|c|}{1.5} & 2\\
\cline{2-9}
 & $\bar{\omega}$ & $\bar{\omega}$-Nb & $\bar{\omega}$-V &$\bar{\omega}$-Nb & $\bar{\omega}$-V & $\bar{\omega}$Nb & $\bar{\omega}$-V & $\bar{\omega}$  \\
 \hline
$A_g$(1)& 882 & 881 & 867 & 878 & 864
& 878 & 865 & 859 \\
\hline
$B_{1g}$(1)& 828 & 826 & 814 & 826 & 814 & * & 808  & 808 \\
\hline
$B_{1g}$(2)& 697 & 697 & 720 & 695 & 717 & 685 & 715  & 705 \\
\hline
$A_{g}$(3)& 530 & 534 & 562 & 541 & 575
 & 549 & 571 & 579 \\
\hline
$A_{g}$(5)& 407 & * & * & * & * & * & * & 406 \\
\hline
$B_{3g}$(7)& 338 &  * & * & * & * & * & * & 333 \\
\hline
$A_{g}$(7)& 298 &  * & * & * & * & * & * & 293 \\
\hline
$A_{g}$(8)& 256 &  * & * & * & * & * & * & 249 \\
\hline
$A_{g}$(9)& 224 &  * & * & * & * & * & * & 220  \\
\hline
$B_{1g}$(11)& 174 & * & * & * & * & * & * & 173  \\
\hline
$A_{g}$(11)& 150 &  * & * & * & * & * & * & 151 \\
\hline
$A_{g}$(12)& 82 &  \multicolumn{2}{|c|}{74} & \multicolumn{2}{|c|}{75} & \multicolumn{2}{|c|}{77} & 81 \\
\hline
$A_{g}$(13)& 68 & \multicolumn{2}{|c|}{59} & \multicolumn{2}{|c|}{59}  & \multicolumn{2}{|c|}{60} & 59 \\
\hline
\end{tabular}
\end{center}
\end{table}

\vspace{3truemm}
In the higher wavenumber region
the most intense peak of each spectrum   is related to the B-O$_t$ bonds and  is observed at a value which is 22 cm$^{-1}$ lower in the spectrum of NiV$_2$O$_6$ than in the spectrum of  NiNb$_2$O$_6$.
According to the Badger's rule, it can be considered that the relative wavenumber of the vibrational modes varies 
with the interatomic distance as 
$\bar{\omega}\propto1/d^{3/2}$.
Since 
the interatomic distance B-O$_t$ is 13\% less when B = V than when B = Nb  (see third column of Table \ref{B_distances}),
the behavior observed in the spectra of our series of samples is opposite to that expected from the Badger's rule. 
A lack of the systematic trend expected from the simple dependence of  $\bar{\omega}$ on the masses and interatomic distances was also
observed in the Raman spectra of the several niobates studied in Ref. \cite{husson_Ag77}; 
there the anomalous variation of the relative wavenumber  was explained by taking into account  the  electronegativity of the M$^{2+}$ cations.
In our case  \mbox{M = Ni$^{2+}$} in both $x=0$ and $x=2$ samples;
notwithstanding, the remaining idea is that
physicochemical factors different from those considered in the simple form of the Badger's rule 
may have to be considered in order to explain the variation in relative wavenumber of the Raman modes from one compound to another.
Below we will present some other examples of anomalous variations of the Raman relative wavenumber with the interatomic distance in different compounds even when the local environment of the ions
remains similar.

\vspace{3truemm}
Table \ref{V5_comparison} shows the position of the most intense peak in the region $\bar{\omega}>700$ cm$^{-1}$ of the Raman spectra in several compounds where the local environment of the V$^{5+}$ cations is octahedral. 
Looking at the set of vanadates AV$_2$O$_6$ (A = Ni, Co, Mg, Zn) with crystalline structure belonging to the \textsl{Pbcn} space group, 
one can see that the monotonic
increase of the relative wavenumber with decreasing V-O$_t$  distance is not verified.
As in Ref. \cite{husson_Ag77}, the higher electronegativity of nickel with respect to magnesium and zinc  can explain the lower relative wavenumber of the $A_g$ Raman mode involving the oxygen atoms in the terminal position even if the V-O$_t$ distance is smaller in the NiV$_2$O$_6$. 
 
\begin{table}
\caption{Structural environment of V$^{5+}$-cations in different compounds and  relative wavenumber of the most intense peak in the region  $\bar{\omega}>700$ cm$^{-1}$ ($\bar{\omega}_{A_{g'}}$). The acronyms are: SG $\equiv$ Space group; CN $\equiv$ coordination number in octahedral (Octa) or square pyramidal (Pyram) 
environment formed by the oxygen atoms;  SS$\equiv$Site symmetry; RP $\equiv$ room-pressure   $^{\ast}$The complete set of results for this compound will be published elsewhere.}\label{V5_comparison}
\begin{center}
\begin{tabular}{|c|c|c|c|c|c|c|c|}
\hline 
Compound & SG & CN & SS & V-O$_t$ \AA & P & $\bar{\omega}_{A_{g'}}$ & Ref.\\
\hline
NiV$_2$O$_6$ & \textsl{Pbcn} & 6 - Octa & 1 & 1.64 & RP & 859
& \\
CoV$_2$O$_6$ & \textsl{Pbcn} & 6 - Octa & 1 & 1.76 & RP & 852
& *\\
MgV$_2$O$_6$ & \textsl{Pbcn} & 6 - Octa & 1 & 1.69 & RP & 916
& \cite{lian_ChinPhysB}\\
MgV$_2$O$_6$ & \textsl{C2/m} & 6 - Octa & m &  1.67 & 4 GPa & 910 
& \cite{tang2016}\\
MgV$_2$O$_6$ & \textsl{C2} & 6 - Octa & 1 & 1.57 & 27 GPa & 
 970 & \cite{tang2016}\\
ZnV$_2$O$_6$ & \textsl{Pbcn} & 6 - Octa & 1 & 1.66 & RP &  918
& \cite{beltran2019}\\
ZnV$_2$O$_6$ & \textsl{C2/m} & 6 - Octa & m & 1.67 & 15 GPa & 945
& \cite{tang2014}\\
ZnV$_2$O$_6$ & \textsl{C2} & 6 - Octa & 1 & 1.66 & 21 GPa & 955 
& \cite{tang2014}\\
$\beta-$V$_2$O$_5$& \textsl{P21/m} & 6 - Octa & m & 1.58/1.65 & RP & 1021/942
& \cite{shvets, grzechnik, hadjean}\\
 $\alpha-$V$_2$O$_5$ & \textsl{Pmmn} & 5+1 - Piram & m & 1.59 & RP & 996
 & \cite{zhou, hadjean, abello}\\
La$_2$LiVO$_6$ & \textsl{Fm3m} & 6 - Octa & & 1.77 & RP & 730
& \cite{choy, demazeau}\\
\hline
\end{tabular}
\end{center}
\end{table}

\vspace{3truemm}
In both MgV$_2$O$_6$ \cite{tang2016} and ZnV$_2$O$_6$ \cite{tang2014} the room pressure  structure (\textsl{C2/m}) decreases its symmetry presenting a phase transformation to a \textsl{C2}-phase 
when pressures of about 20 GPa and 17 GPa, respectively, are applied. 
The relative wavenumber of the most intense peak of the Raman spectrum of MgV$_2$O$_6$ ($\bar{\omega}=922$ cm$^{-1}$) reduces  when external pressure up to P = 3.9 GPa is applied, there an increasing in the coordination number (CN) from 5+1 to 6 occurs \cite{tang2016}.
At higher pressures the relative wavenumber of the Raman mode
increases continuously until at least P = 28 GPa.
In both the Mg and Zn cases the most intense peak of the Raman spectrum 
increases in relative wavenumber as the pressure increases.
Simultaneous structural studies show that this shift in the Raman relative wavenumber is accompanied by an increasing distortion of the AO$_6$ and VO$_6$ octahedra. 
As can be seen in Table \ref{V5_comparison}, in the case of MgV$_2$O$_6$ the reduction  of symmetry between \textsl{C2/m} and  \textsl{C2}-phases is accompanied by a decrease in the V-O distance and 
an increase in the relative wavenumber of the Raman mode.
However, in the ZnV$_2$O$_6$ the CN, the octahedral environment and the V-O distance in the three structures (\textsl{Pbcn, C2/m} and \textsl{C2}) remains almost unaltered but the relative wavenumber of the Raman mode varies greatly and it is the highest for the less symmetric crystalline structure.

%
%
\vspace{3truemm}
The case of V$_2$O$_5$ is interesting because the application of external pressure induces not only a phase transition to a crystalline phase of lower symmetry, but it also changes the local environment of the V$^{5+}$ cation.
The $\alpha-$V$_2$O$_5$, stable at room conditions, is formed by edge-sharing bi-pyramidal V$_2$O$_4$ chains where all the V-atoms have equivalent crystallographic positions \cite{shvets, abello}.
When pressures higher than P = 4.5 GPa are applied, an increasing of the CN  from 5+1 to 6 accompanied of a
structural phase transition from \textsl{Pmmn}  to \textsl{P2$_\textrm{1}$/m} space group is induced. 
In the high-pressure phase, $\beta$-V$_2$O$_5$, the crystallographic sites of V, O1, and O2 atoms split in two non equivalent positions which form two different octahedral environments for the V$^{5+}$ cations \cite{ shvets, abello, hadjean, grzechnik}.  
In consequence, not one but two $A_g$ peaks are observed in the Raman spectrum of the $\beta$-V$_2$O$_5$;
one of those peaks is ascribed to one  bond with interatomic V-O distance which is almost the same
than that in the $\alpha-$phase (1.59 \AA);
however, an increase of 25 cm$^{-1}$ is observed in the relative wavenumber of the mode of the $\beta-$phase with respect to the $\alpha-$one \cite{hadjean}. 
One can summarize these experimental observations by saying that the reduced symmetry of the overall structure and increasing CN
favored an increasing in the relative wavenumber of the $A_g$ mode in the $\beta$-V$_2$O$_5$
with respect to that of the $\alpha$-V$_2$O$_5$.

\vspace{3truemm}
If we compare now the  \textsl{Pbcn}-NiV$_2$O$_6$ compound 
with the double perovskite La$_2$LiVO$_6$ where the octahedra are rather regular \cite{choy},
a strong contrast appears.
In the \textsl{Pbcn}-NiV$_2$O$_6$ the VO$_6$ octahedra are  strongly distorted, so these two compounds can be taken as the extreme behavior of the V$^{5+}$ state octahedral environment.
In Table \ref{V5_comparison} it is clear that the
La$_2$LiVO$_6$ shows by far the lowest relative wavenumber of the higher intensity peak, while the V-O distance is the same that in the \textsl{Pbcn}-CoV$_2$O$_6$.
These facts, and those presented in the previous paragraphs point to a general conclusion in terms of
the symmetry not only of the local environment but also of the overall crystalline structure where the V$^{5+}$ ions are inserted. 
In general, the relative wavenumber of the highest intensity Raman peak is very sensitive to the specific details of the crystalline structure as a whole 
and so it behaves almost as a fingerprint of the compound; apparently this mode tends to appear at lower frequencies in more symmetric space groups and higher coordination numbers of the B-atoms.

\vspace{3truemm}
Returning to the Raman spectra of the samples  NiNb$_2$O$_6$ and NiV$_2$O$_6$ studied here, 
we will point to some differences both  in the local and global
environments of the V$^{5+}$ and Nb$^{5+}$ cations.
From the variation of the angles B-O-B, $\theta_1$ and $\theta_2$, and from the calculated dispersion  of the B-O distances,
we concluded that the VO$_6$ octahedra are more
distorted than the NbO$_6$ ones. 
The mean distance B-O is more than 7\% less
when B = V than when B = Nb,
but it does not necessarily lead  to a higher
stiffness of the B-O bonding. 
The lower B-O distance when B = V can be barely due  to the fact that the effective radius of the V$^{5+}$ ($r=0.59$ \AA) is smaller than the effective radius of Nb$^{5+}$ ($r=0.69$) \cite{shannon}.
In the counterpart, the higher effective nuclear charge of the Nb$^{5+}$ ($Z_{\textrm{eff}}=13.25$)  with respect to that of the V$^{5+}$
($Z_{\textrm{eff}}=11.8$) \cite{demazeau}
could lead to higher force constants of the bonds even if the oxygen atoms are further in the former case.
Additionally, we showed the previous paragraphs  that the relative wavenumber of the most intense Raman peak is affected by the whole crystalline structure and not only by the local environment of the B-cations. 
If we look now to the nickel ion environment, the 
distortion calculated for the NiO$_6$ octahedra in the $x=0$ sample is $\Delta=6.89\times10^{-4}$,  against $\Delta= 1.27\times10^{-4}$ in the $x=2$ sample. 
Thus we conclude that, opposite to the situation of the B$^{5+}$-cations, the environment of the Ni$^{2+}$- cations
is much more symmetric in the NiV$_2$O$_6$ than in the NiNb$_2$O$_6$. 
Then, it is a possibility that the force constant of the Ni-O bond is higher in the NiV$_2$O$_6$ than in the NiNb$_2$O$_6$ and that, similarly to the effect of the electronegativity, 
this weakens the force constant of the V-O bonds thus reducing the relative wavenumber of the most intense Raman peak with respect to that observed in
the NiNb$_2$O$_6$ Raman spectrum.

\vspace{3truemm}
Until now we have discussed mainly over the Raman spectra of the samples  $x=0$ and $x=2$. 
We would like to focus now  on the Raman spectra of the samples
with intermediate quantities of niobium and vanadium, $0<x<2$. 
A common characteristic to all these samples is that their Raman spectra are mainly composed of very spread intensity distributions  which correspond to obvious increasing from the background line,
but where individual peaks can not be separated.
It was said above that one part of the widening 
of the peaks seems to be intrinsic to the 
highly distorted VO$_6$ octahedra as already seen
in pure NiV$_2$O$_6$. 
However, the  absence of a continuous evolution of the modes in the Raman spectrum with increasing vanadium content makes us think of a two-modes
behavior as another contribution to the widening of the Raman  peaks in the samples
with both NbO$_6$ and VO$_6$  pseudo-octahedra.

\vspace{3truemm}
The two modes evolution of the Raman spectra in a solid solution is well described in the study done  for semiconductor alloys in Ref. \cite{pages_clusters}. 
Here, we can think of our samples of NiNb$_{2-x}$V$_x$O$_6$ as being formed by an increasing quantity of V-atoms getting forced into a NiNb$_2$O$_6$ matrix;
the fact that the Raman spectrum evolves showing a two modes behavior along the series of samples means that the V-atoms do not spread homogeneously in the unitary cells of the matrix,
and so one can not think of the
existence of  a virtual B-atom possessing intermediate properties between those of the niobium and those of the vanadium,
but instead clusters of Nb and V-octahedra whose
size evolves with the value of $x$ along the solid solution.
Due to the significant differences we saw in the NbO$_6$ and VO$_6$  structural blocks, 
the tendency towards clustering similar NbO$_6$ or else VO$_6$ octahedra is actually expected. 
In our samples, the widening of the Raman peaks  due to cluster formations comes from the independent vibrations of the local bonds in the Nb and V-octahedra;
these vibrations manifest separately and, 
as a consequence the Raman peaks do not evolve so much in relative wavenumber but in intensity depending on  the increase or decrease of the  quantity of one determined atomic species.
In general the clustering reduces the grain-size of each individual structure thus broadening the overall Raman signature.
This reduction of the crystallite size also manifest as a broadening of the peaks on the XRD patterns of samples containing both Nb and V atoms
with respect to those of the samples containing only Nb or only V; 
the FWHM of the plus intense Bragg peak (311) goes from 0.14 and 0.11 in $2\theta$for  NiNb$_2$O$_6$ and NiV$_2$O$_6$, respectively,
to a mean of 0.22 in the other samples (see Supporting Information file).
Here we may think that the segregation in clusters of the V-atoms in  the very stable NiNb$_2$O$_6$ matrix is one possible explanation for the difficulties faced  to obtain the pure $x=0.5$ sample.
At this point it is worth to recall that Raman spectroscopy is a probe very sensitive to the local atomic environment, whereas the crystal structure analysis performed by XRD provides a complementary information averaging over several unit cells leading to mean interatomic distances.
This averaging produces the sensation of having one single structural phase 
even if in some cases it is also imposible to separate the individual Bragg reflections in the XRD patterns of the samples with intermediate compositions of Nb and V;
thus, the observed continuous evolution of the XRD pattern and the unit-cell parameters 
may not describe the real evolution of the local atomic environments.

\begin{figure}
\centering
\includegraphics[keepaspectratio,width =10truecm]{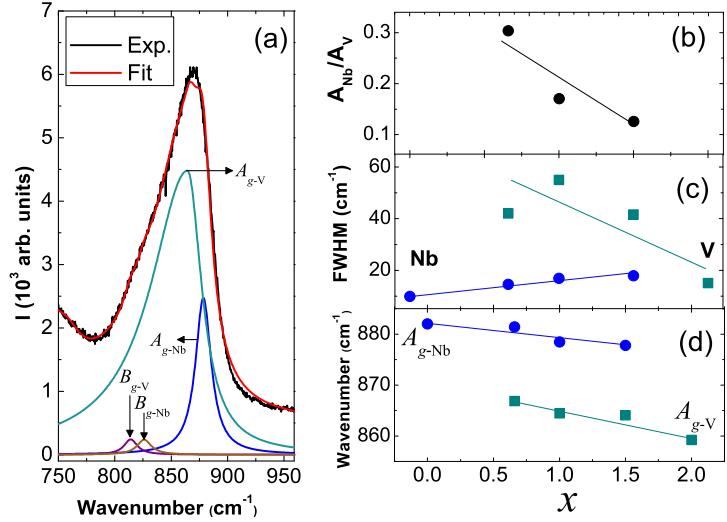}
 \caption{Two-modes behavior exemplified in the most intense peak $A_g$(1) of the Raman spectra of the NiNb$_{2-x}$V$_x$O$_6$ series of samples. (a)Fitting of the highest Raman peak of sample NiNbVO$_6$, the fitting curve is the sum of  peaks $A_g$(1) and $B_g$(1) of samples $x=0$ and $x=2$; (b)Ratio of the areas of peaks $A_g$(1) for samples $x=0$ (A$_\textrm{Nb}$) and $x=2$ (A$_\textrm{V}$); (c)FWHM of the  $A_g$(1) peaks as a function of the vanadium content; (d)Variation in relative wavenumber of the peak $A_g$(1) of Nb and V-octahedra with vanadium content. Solid lines are guides for the eyes.} \label{two_modes}
\end{figure}

\vspace{3truemm}
Because of the atomic vibrations in the NbO$_6$ and VO$_6$ octahedra manifest separately in the samples with intermediate composition of Nb and V,
their experimental Raman spectra can be fitted by adding the separated contributions of peaks for the pure Nb and pure V samples. 
An illustration of how the two-modes behavior manifest in the most intense peak of the Raman spectra of our
NiNb$_{2-x}$V$_x$O$_6$ series of samples is shown in Fig. \ref{two_modes}.
For the sample $x=1$ it is exemplified in Fig. \ref{two_modes}(a) that the form of the peak can be reproduced by joining the  corresponding peaks to the Raman spectra of the samples with  extreme compositions $x=0$ and $x=2$.
The broadening that appears at the low wavenumber side is reproduced
by the appearing of the  peaks $B_g$(1) (this is signaled by the symbol $\bullet$ in Fig. \ref{raman_specta} for sample $x=0$) and by taking into account an asymmetry of the peak $A_g$(1)  of sample $x=2$.
This asymmetry is possibly due to the same type of local disorder mentioned above to explain the widening of the Raman peaks of the NiV$_2$O$_6$ sample, and it  is probably increased by the presence of  octahedra with different sizes and in different exterior environments. 
The increasing of local disorder also manifests  
as a general enlargement of the FWHM of the individual peaks necessary to 
make the fitting of the Raman peaks of  samples with intermediate compositions of Nb and V (see panel (c) of Fig. \ref{two_modes}).
The reduction of the contribution of the NbO$_6$ octahedra to the Raman spectrum as the V atoms are substituted for Nb ones is seen in the reduction of the ratio of the areas of individual peaks,
A$_\textrm{Nb}$/A$_\textrm{V}$, with increasing $x$ (Fig. \ref{two_modes}(b)).
Finally, the positions  of the individual $A_g$(1) peaks attributed to the Nb and V octahedra, which are used to fit the $A_g$(1) peaks of the intermediate composition samples   along the whole series, is plotted as a function of $x$ in Fig. \ref{two_modes}(d) and their values are
presented in the third row of Table \ref{Raman_peaks}.
The results of the fits performed in the region $500<\bar{\omega}<1000$ of the Raman spectra for all samples are shown in the Supporting Information;
the peaks that could be most clearly separated are also reported in columns 3 - 8 of Table \ref{Raman_peaks}.

\begin{figure}[ht] 
\centering
\includegraphics[keepaspectratio,width =10truecm]{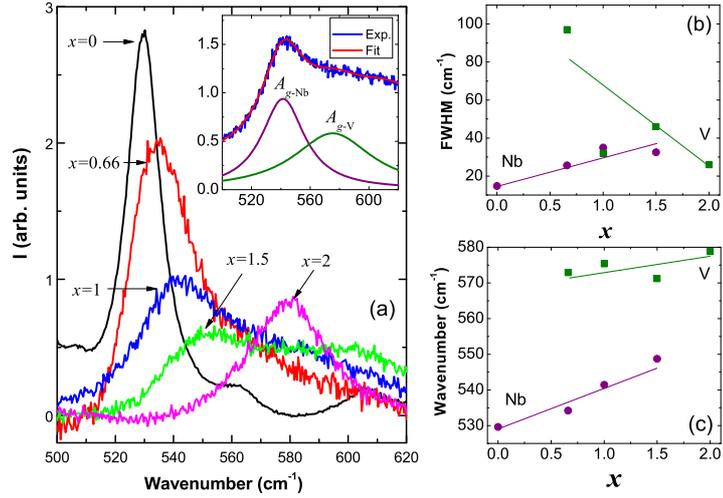}
 \caption{(a) Evolution of the Raman-mode $A_g$(3) with the concentration of vanadium along the NiNb$_{2-x}$V$_x$O$_6$ series of samples; in order to improve the visualization, the data for $x=0$ is multiplied by 0.5. The inset shows the experimental curve of the sample $x=1$ fitted by the two-modes of the extreme compositions; (b)FWHM of the  peaks used to perform the fitting of the $A_g$(3) mode as a function of the vanadium content; (d)Variation in relative wavenumber of the center of the peaks just described along the whole series of samples. Solid lines are guides for the eyes.} \label{two_modes_Ag2}
\end{figure}

\vspace{3truemm}
We will consider now the second region
of frequencies ($500<\bar{\omega}<800$) cm$^{-1}$ in the Raman spectra  of our NiNb$_{2-x}$V$_x$O$_6$ series of samples.
Disregarding the peak related to the mode $B_{1g}$, which is too wide or too weak to make
any trustable analysis over it, the only peak remaining in this region is the one related to the mode $A_{g}$(3) which describes the B-O$_b$ stretching vibrations. 
Looking only at the two extreme compositions $x=0$ and $x=2$, the distances B-O$_b$
are sensibly smaller when B = V than when B = Nb.
Based on this statement, it is expected from the Badger's rule that the force constants of the V-O$_b$ bonds are stronger than those of the Nb-O$_b$ ones, and so 
the modes related to these bonds are  expected to appear at higher frequencies in the NiV$_2$O$_6$ than in the NiNb$_2$O$_6$, as it is experimentally observed.
Regarding now to the complete set of samples, one can track the monotonic increasing in the relative wavenumber where the maximum of the peak $A_{g}$(3) is observed in Fig. \ref{two_modes_Ag2}; it varies from $\bar{\omega}=530$ cm$^{-1}$ for $x=0$ to \mbox{$\bar{\omega}=580$ cm$^{-1}$} for $x=2$.
However, as it can be verified in Table \ref{B_distances}, the same monotonic behavior is not observed in any of the distances ascribed to the oxygen atoms in the bridge positions.
In consequence, the linear behavior expected from the Badger's rule in the graphics
$\bar{\omega}^2$ vs. $1/d^3_{\textrm{B-O}_b}$ is not verified (not shown).
We should notice that in the frame of the two modes behavior, a possible reason for this is that the values calculated by the XRD represent mean B-O$_b$  distances 
which may not correspond to a good representation of the real local environments of the Nb and V atoms.
As it was discussed above, it is probable that because of the very different octahedral environments of the Nb and V atoms,
the samples with $0<x<2$ are formed of 
clusters of separated VO$_6$ and NbO$_6$ octahedra which can not be represented by one virtual medium atom as the one considered by the XRD technique.
Here we invoke again the  two modes behavior in order to explain the evolution of the Raman peaks along our series of samples in this range of wavenumbers;
it is exemplified for the  mode $A_g$(3) in the inset of Fig. \ref{two_modes_Ag2} (a).
Because of the contribution of neighbor peaks which are wider and closer in this region of the spectrum than in the higher relative wavenumbers, the fitting 
is more difficult and some shift in the center of the peaks attributed to the pure Nb and pure V-octahedra is even found as it can be seen in the panel (c) of Fig. \ref{two_modes_Ag2}
and in row six of Table \ref{Raman_peaks}.  

\vspace{3truemm}   
In the third region of the spectra ($\bar{\omega}<500$ cm $^{-1}$), following the different Raman modes across the solid solution is difficult because of a general broadening of the peaks in the samples with $x>0$. 
The  modes $A_g$(12) and $A_g$(13) which appear respectively at $\bar{\omega}=82$ cm$^{-1}$ and $\bar{\omega}=68$ cm$^{-1}$ in the spectrum of sample $x=0$ show an abrupt shift to lower wavenumbers when $x$ goes to  $0.66$. 
With further increasing proportion of V in the solid solution, the mode $A_g$(12) increases regularly in wavenumber to reach the same position in NiV$_2$O$_6$ as in NiNb$_2$O$_6$; whereas the mode $A_g$(13) remains at a constant wavenumber of 60 cm$^{-1}$ for all samples with $x\geq0.66$. 
The modes $A_g$(12) and $A_g$(13) are controlled by the O-Nb-O/Ni-O-Nb bending and by the Ni-O stretching, respectively.
The abrupt blue shift detected in the spectra of samples with $x>0$ is in agreement with the increase of the Ni-O distances as  deduced from the XRD results (see Table \ref{Ni_distances}).

\section{Conclusions}

\vspace{3truemm}
The compound NiV$_2$O$_6$ with the unusual columbite-type structure was successfully synthesized under conditions of high pressure and temperature.
The  structural characterization shows that this material is formed by two fundamental structural units of octahedral symmetry, one very regular around the Ni-atoms and one very distorted around the V-ones.
While conserving the \textsl{Pbcn} structure, the gradual substitution of vanadium for niobium in the 
NiNb$_{2-x}$V$_x$O$_6$ series of compounds
produces the anisotropic shrink of the unit-cell parameters:
between the two extreme compositions the variation in the  $a$ and $c$ parameters is almost the same but twice that observed in the $b$ parameter.
When compared, the BO$_6$ octahedral units of the NiB$_2$O$_6$-\textsl{Pbcn} compound are smaller but more distorted for B = V than for B = Nb;
in contrast, the NiO$_6$ structural units are bigger and more symmetric in the former case.

We report the first Raman spectrum of the NiV$_2$O$_6$-\textsl{Pbcn} HPHT
polymorph and extended the list of the previous observed modes for the
NiNb$_2$O$_6$.
The Raman spectrum of NiV$_2$O$_6$-\textsl{Pbcn} is characterized by relatively wide peaks. 
This widening was ascribed to the very distorted octahedron.
The evolution of the Raman spectrum with increasing vanadium content along the series of samples NiNb$_{2-x}$V$_x$O$_6$ was analyzed by considering a two-modes behavior consisting in the formation of clusters of separated  vanadium and niobium octahedra.
The Raman spectra of  the samples  with intermediate compositions of Nb and V ($0<x<2$) are formed by a set of entangled peaks forming very broad structures all along the relative wavenumber ranges. 
These spectra contain the modes of  vibration related to four different structural environments: NbO$_6$ octahedra, very distorted VO$_6$ octahedra and two  NiO$_6$ octahedra with different sizes and distortion indexes. 
The enlargement of the Raman peaks of the samples with intermediate compositions of Nb and V
shows that there is an inhomogeneous distribution of Nb and V-octahedra leading to structural disorder and atomic segregation.
The comparison of the higher wavenumbers region of the Raman spectra of NiNb$_2$O$_6$ and NiV$_2$O$_6$
led us to conclude that
the smaller size of  V$^{5+}$,
which favors an increasing symmetry of the structural arrangement of oxygen around the central Ni atom, boosts the  ionic character of the  Ni-O bonds  in the samples rich in vanadium.
Despite of  the lower  V-O distances when compared to the Nb-O ones, the increasing ionic character of the  Ni-O bonds in samples containing vanadium results in a lower force constant of the V-O bonds in the VO$_6$ octahedra with respect to that of the Nb-O bonds in the NbO$_6$ octahedra.
Because of the changes induced in the electronic degrees of freedom produced in the magnetic atoms by the reinforcement of their atomic bonds, important differences in the behavior of the magnetic properties of the NiV$_2$O$_6$ with respect to the NiNb$_2$O$_6$ can be preview.
The shifts in the force constants of the B-O bonds can also give important differences in the optical and mechanical properties of the compounds all along the series of samples studied here and the quantity of vanadium could be used as a tuning parameter to suppress or reinforce such physical characteristics.

Further theoretical studies may be necessaries in order to completely understand the observed changes in the Raman spectrum along the series of samples NiNb$_{2-x}$V$_x$O$_6$ studied here.

\section{Supporting Information}

Complete list of experimentally identified Raman modes for NiNb$_2$O$_6$.
Fitting curves and tables of the Raman modes identified in the spectrum of each sample in the region 500 - 1000 cm$^{-1}$.

\section{Aknowledgments}

This work was financed by the french-brazilian cooperation program  'CAPES-COFECUB', process 88887.321681/2019-00,
and the french institution 'Centre National de la Recherche Scientifique', CNRS. 
We thank to the professional team of the X-Press pole of the 'N\'eel Institut', specially Celine Goujon and Murielle Legendre for the technical support in the preparation of the samples.

\bibliographystyle{ieeetr}
\bibliography{Raman_Ni_article_bibliography}

\end{document}